\documentstyle[12pt]{article}
\title{\baselineskip=9mm
Fusion barrier distributions \\ in systems with finite excitation energy}
\author{K. Hagino,$^{1,3}$ N. Takigawa,$^{1,3}$ and A.B. Balantekin$^{2,3}$
\\ \\
\medskip
{\it $^1$ Department of Physics,
Tohoku University, Sendai 980--77, Japan}\\
{\it $^2$ Physics Department, University of Wisconsin, }\\
{\it Madison, Wisconsin~53706}\\
{\it $^3$ Institute for Nuclear Theory, University of Washington, }\\
{\it Seattle, Washington 98915 }\\
}
\date{}

\begin{document}
\maketitle

\begin{center}
{\bf Abstract}
\end{center}

Eigen-channel approach to heavy-ion fusion reactions 
is exact only when the excitation energy of the intrinsic 
motion is zero. In order to take into account effects of 
finite excitation 
energy, we introduce an energy dependence to weight factors 
in the eigen-channel approximation. 
Using two channel problem, we show that 
the weight factors are slowly changing functions of 
incident energy. This suggests that the concept of the fusion barrier 
distribution still holds to a good approximation even when the excitation 
energy of the intrinsic motion is finite. 
A transition to the adiabatic tunneling, where the coupling 
leads to a static potential renormalization, is also discussed. 

\medskip

\noindent
PACS number(s):
25.70.Jj, 24.10.Eq

\newpage

It has been well recognized that cross sections of heavy-ion 
fusion reactions at energies near and below the Coulomb barrier 
are strongly influenced by coupling of the relative motion of 
the colliding nuclei to nuclear intrinsic motions\cite{BT97}. 
When the intrinsic degree of freedom has a degenerate spectrum, 
fusion cross sections 
can be calculated in the sudden approximation. 
In this limit, the coupling gives rise to a distribution of 
fusion potentials. The fusion cross section is then given 
by an average over the contributions from each fusion barrier 
with appropriate weight factors\cite{HTBB95,NBT86,E81}. 
Based on this idea, a method was proposed to extract 
the barrier distribution 
directly from the fusion excitation function using 
the second derivative with respect to 
the product of the fusion cross section 
and the center of mass energy, $E \sigma$ \cite{RSS91}. 
This stimulated precise measurements of the fusion cross sections for 
several systems\cite{LDH95}. 
The extracted barrier distributions have then 
been found to be very sensitive to the structure 
of the colliding nuclei, e.g. the sign of hexadecapole deformation 
parameter, while the fusion cross section itself is rather 
featureless\cite{LDH95,LLW93}. 

In a rigorous theoretical interpretation, the barrier distribution 
representation, i.e. the second derivative of $E\sigma$, has a clear 
physical meaning only if the excitation energy of the intrinsic motion 
is zero. 
Nonetheless, this analysis has been successfully applied to 
systems with relatively 
large excitation energies\cite{LDH95,MDH94,SACN95}. 
For example, the second derivative of $E\sigma$ for $^{16}$O + $^{144}$Sm 
fusion reaction 
clearly shows the effects of coupling to the octupole phonon state in 
$^{144}$Sm, whose excitation energy is 1.8 MeV, much more clearly 
than the 
fusion cross section itself\cite{LDH95,MDH94}. Also 
the analysis of the fusion reaction between $^{58}$Ni and $^{60}$Ni, 
where the excitation energies of quadrupole phonon states 
are 1.45 and 1.33 MeV, 
respectively, shows that the barrier distribution 
representation depends strongly on the number of phonon states 
included in coupled-channels calculations\cite{SACN95}. 
These analyses suggest that 
the representation of fusion process in terms of the second derivative of 
$E\sigma$ 
is a very powerful method to study the details of the effects of 
nuclear structure, 
irrespective of the excitation energy of the intrinsic motion. 

Despite these successes, there remains the question 
whether the second derivative of $E\sigma$ represents a 
\lq\lq distribution \rq\rq of fusion potential barriers when the 
excitation energy of the intrinsic motion is not zero. 
To address this problem, one may attempt to use the 
constant coupling approximation, where the coupling form factor is 
assumed to be constant throughout the interaction range \cite{DLW83}. 
In this approximation, 
the unitary transformation gives the fusion cross section 
as a sum of eigen-channel cross sections. 
The constant coupling approximation has, however, a serious drawback 
in that the results strongly depend on the position where the strength 
of the coupling Hamiltonian is evaluated. 
Therefore, one needs to use a different approach to 
address the question. 

In this paper we treat the weight factors as energy 
dependent variables. 
The possibility of the energy dependence of weight factors 
when the excitation energy is finite has been suggested in ref. \cite{RD93}. 
Here, we explicitly study the energy dependence by performing 
exact coupled-channels calculations. 
It will be shown that the energy dependence is quite weak 
for a wide range of excitation energy, suggesting that 
the eigen-channel approximation works well even when the excitation 
energy of the intrinsic motion is not small. 

Let us consider, for example, the case where 
the intrinsic degree of freedom has 
only two levels. The coupled-channels equations then read
\begin{equation}
\left[-\frac{\hbar^2}{2\mu}\frac{d^2}{dR^2}+U(R)
+\left(
\begin{array}{cc}
0&F(R)\\ F(R)&\epsilon
\end{array}
\right)\right]
\left(
\begin{array}{c}
u_0(R)\\ u_1(R)
\end{array}
\right)
=E
\left(
\begin{array}{c}
u_0(R)\\ u_1(R)
\end{array}
\right),
\end{equation}
where $U(R)$ and $\mu$ are the bare potential and the reduced 
mass of the 
relative motion, respectively. $F(R)$ is the coupling form factor 
and $\epsilon$ is the excitation energy of the intrinsic motion, 
respectively. 
When the excitation energy $\epsilon$ is zero, the 
unitary matrix which diagonalizes the coupling matrix 
is independent of the position $R$, and the barrier penetrability 
is exactly given by
\begin{equation}
P(E)=\frac{1}{2}\left[P_0(E; U(R)+F(R)) + P_0(E; U(R)-F(R))\right], 
\end{equation}
where $P_0(E; V(R))$ is the penetrability of potential $V(R)$ 
at energy $E$. 
In this limit, the original single barrier splits into two 
effective potential barrier with equal weights. 

If the excitation energy $\epsilon$ is not zero 
the unitary matrix depends explicitly on the position $R$. 
The unitary matrix, therefore, 
does not commute with the kinetic energy operator, 
and hence simple eigne-channel approach does not hold. 
An approximation which is often made is to assume that the 
coupling form factor $F(R)$ is a constant. In this approximation, 
the unitary matrix becomes independent of $R$, and the 
penetrability is given by a formula similar to eq. (2) but with 
different eigen-values and weight factors \cite{DLW83}. 
In order to take into account the radial dependence of the 
coupling form factor, an approximation was proposed, 
where the coupling matrix is diagonalized to obtain the 
eigen-barriers at each position 
of the relative motion $R$ while the weight factors are 
calculated at a chosen position $R_w$ \cite{DL87}. 
The computer code CCMOD uses this approximation\cite{DNA92}. 

Even when we introduce this approximation, the weight factors 
are still functions of the chosen position $R_w$, and the results 
might strongly depend on that choice. 
Usually $R_w$ is chosen to be the 
position of maximum of the bare potential barrier\cite{DL87,DNA92}. 
However, there is no theoretical justification that this is the 
optimum choice, and 
furthermore, it is not obvious whether one can determine the 
weight factors independent of the incident energy. 

In order to avoid these drawbacks and examine the energy dependence 
of the weight factors, we parametrize the penetrability as 
\begin{equation}
P(E)=v_+(E)P_+(E) + v_-(E)P_-(E),
\end{equation}
and evaluate the weight factors $v_{\pm}$ at each incident energy $E$. 
Here $P_{\pm}(E)$ are the penetrabilities of the eigen-potentials
$U(R)+\lambda_{\pm}(R)$, respectively, $\lambda_{\pm}(R)$ being 
the eigen-values of the coupling matrix at each position of $R$, which 
are given by 
\begin{equation}
\lambda_{\pm}(R)=\left(\epsilon \pm \sqrt{\epsilon^2+4F(R)^2}\right)/2. 
\end{equation}
If the weight factors slowly vary as functions of the incident 
energy, the effects of channel coupling can be interpreted in terms 
of the barrier {\it distribution}, even when the excitation energy 
is non-zero. 
In the appendix, we generalize eq. (3) to multi-channel case 
using the path integral approach. 
Since the weight factors have to satisfy the unitarity 
condition\cite{NBT86}, they can be uniquely determined in the present 
two level problem, and are given by 
\begin{equation}
v_+(E)=(P(E)-P_-(E))/(P_+(E)-P_-(E)),
~~~~v_-(E)=(P_+(E)-P(E))/(P_+(E)-P_-(E)).
\end{equation}

We performed coupled-channels calculations to examine the 
energy dependence of the weight factors. 
To this end, the coupled-channels equations have to be solved 
with good accuracy, since the penetrabilities $P(E)$ and 
$P_{\pm}(E)$ in eq. (5) are exponentially small quantities 
at energies below the barrier. 
The incoming wave boundary condition, which is often used in 
coupled channels calculations for heavy-ion 
collisions \cite{LP84}, can bring some numerical errors, 
though they would be small enough for the purpose of calculating 
fusion cross sections. 
In order to avoid this, we use here a schematic model of heavy-ion 
fusion reactions by Dasso {\it et al.}\cite{DLW83}, where 
the radial motion between colliding nuclei is treated 
as a one dimensional barrier penetration problem. 
Following ref. \cite{DLW83}, we assume the gaussian 
shape for both the bare potential and the coupling form factor,
\begin{equation}
U(R)=U_0e^{-R^2/2s^2}, ~~~F(R)=F_0e^{-R^2/2s_f^2}. 
\end{equation}
The parameters are chosen following ref. \cite{DLW83} to be 
$U_0$ = 100 MeV, $F_0$ = 3 MeV, and $s = s_f$ = 3 fm, respectively, 
which mimic the fusion reaction between two $^{58}$Ni nuclei. 
We have checked that our conclusions do not significantly change as 
long as the value of $s_f$ is not too small. 
The mass $\mu$ and the excitation energy $\epsilon$ are taken 
to be 29 $m_N$, $m_N$ being the nucleon mass, and 2 MeV, respectively. 

The upper panel of Fig. 1 shows potential barriers for the 
present problem. The dotted line is the bare potential $U(R)$ and 
the solid and the dashed lines are the eigen-potentials 
$U(R)+\lambda_-(R)$ and $U(R)+\lambda_+(R)$, respectively. 
If we adopt the eigen-channel picture, Fig. 1 means that the potential 
barriers are \lq\lq distributed \rq\rq 
and the original bare potential (the dotted line) splits into 
the two potentials (the solid and the dashed lines). 
Because of the non-commutatibity of the unitary matrix which 
diagonalizes the coupling matrix, and the kinetic energy operator, 
the standard eigen-channel picture does not apply. 
If we ignore this non-commutatibity (i.e., if we adopt with the 
CCMOD prescription) the weight factors are given by 
\begin{equation}
w_{\pm}(R)=F(R)^2/\left(F(R)^2+\lambda_{\pm}(R)^2\right). 
\end{equation}
They are shown in the lower panel of Fig. 1 as a function of the distance 
$R$. 
Although they do not vary so much near the barrier position, 
their changes are appreciable throughout the barrier 
region. Therefore, the prescription to fix the weight factors 
to the values at the barrier position may not be satisfactory. 

The results of exact coupled-channels calculations 
are shown in Fig. 2. The first and the second 
panels are the penetrabilities 
and their first derivative with respect to the energy, respectively. 
The latter corresponds to the second derivative of 
$E\sigma$ \cite{RSS91}. We use the point difference formula with 
$\Delta E$ = 2 MeV to obtain the first derivative of the 
penetrability, as is often done in the analyses of heavy-ion 
fusion reactions\cite{LDH95}. 
The first derivative of the penetrability has a clear double-peaked 
structure, which could be associated with the two eigen-potential 
barriers and could thus be interpreted in terms of a \lq barrier 
distribution \rq. In order to see whether this is the case, 
we plot in 
the last panel of Fig. 2 the optimum weight factors defined by 
eq. (5). The solid and the dashed lines in the 
figure correspond to the weight factors for the lower and the higher 
potentials, respectively. 
We observe that the optimum weight factors change only slightly 
as functions of the incident energy; their change is barely 2.6 \% 
from 10 MeV below the barrier to 10 MeV above the barrier. 
Another important result of this calculation is that the weight 
factors are considerably different from those estimated in the usual 
way, i.e. at the barrier position. 
When the constant coupling approximation was first introduced, it was 
expected that the choice of the position where the weight factors are 
estimated is not critical to calculate tunneling probabilities because 
the weight factors vary slowly across the barrier region\cite{DLW83,DL87}. 
The position was then chosen to be the barrier position of the 
uncoupled barrier. 
Contrary to this expectation, our calculations show that 
determining the weight factors at the barrier 
position does not give proper weight factors even though the weight factors 
do not have a strong radial dependence. 
The situation would be much more serious in realistic calculations, since 
the weights are still changing at the barrier position due to the 
fact that the coupling extends outside the Coulomb barrier\cite{RD93}. 

The same calculations were repeated for different values of the 
excitation energy $\epsilon$ and the results are shown in Fig. 3. 
The quantities shown in each panel are the same as 
those in Fig. 2 except for the third panel, where only the optimum 
weight factors for the lower barrier are plotted. 
We again observe that the weight factor changes only
marginally as a function of the incident energy, even when 
the excitation energy is finite. 
We thus conclude that the eigen-channel approach is still applicable 
even when the excitation energy of the intrinsic motion is finite. 

When the excitation energy $\epsilon$ is zero, 
the barrier distribution, i.e. the first derivative of the 
penetrability has two symmetric peaks and 
the weight factor has no incident energy dependence (the solid line). 
As the excitation energy increases, some strength is transfered from 
the higher peak to the lower peak, and the barrier distribution 
becomes asymmetric. 
As the excitation energy significantly exceeds the curvature 
of the bare barrier, which is about 4 MeV in our example, 
one expects to reach the adiabatic limit. 
To illustrate this, figure 4 shows the influence of the coupling to 
an excited state whose energy is 8 MeV. 
The figure also contains the result for the 
the no coupling case (the dotted line) for comparison. 
In this case, the weight for the higher peak is considerably 
smaller than that for the lower peak, and the barrier 
distribution has essentially only a single peak. 
The peak position, however, is shifted towards a lower 
energy. This is consistent with the adiabatic picture, i.e. the  
main effect of the coupling to a state whose excitation energy is 
much larger the barrier curvature is to introduce 
an energy independent shift of the potential accompanied with the mass 
renormalization\cite{THA95,THAB94}. 
Correspondingly, the barrier distribution is only shifted 
without significantly changing its shape unless the coupling 
form factor has a strong radial dependence\cite{HTDDL97}. 
It is thus clear that 
the second derivative of $E\sigma$ can represent a wide
range of situations between the two extreme limits, i.e. 
from the adiabatic limit where the coupling leads to an adiabatic 
potential renormalization, to the sudden 
limit where the coupling gives rise to a barrier distribution. 

In summary, we have discussed the energy dependence of the 
weight factors in the eigen channels approximation. 
Performing exact coupled-channels calculations, we showed that the 
energy dependence is very weak, regardless of the excitation energy 
of the intrinsic motion. 
Therefore, it makes sense to call the second derivative of 
$E\sigma$ as the fusion barrier distribution, even when the 
excitation energy is finite. 
We also discussed a transition from the sudden to 
the adiabatic tunneling limits, and showed that the fusion barrier
distribution can represent these two limits in a natural way. 

\bigskip

The authors thank M. Dasgupta for useful discussions.
They also thank the Institute for Nuclear Theory (INT) 
at the University of Washington for its 
hospitality and the Department of Energy for 
partial support during the course of this work. 
The work of K.H. was supported by the Japan Society for the Promotion 
of Science Research Fellowships for Young Scientists.
This work was supported by 
Monbusho International Scientific Research Program: 
Joint Research, Contract No. 09044051 and 
the Grant-in-Aid for General Scientific Research,
Contract No.08640380, 
from the Japanese Ministry of Education, Science and Culture, 
and by U.S. National Science Foundation Grant No. PHY-9605140. 

\bigskip

\begin{center}
{\bf APPENDIX: EIGEN-CHANNEL APPROXIMATION IN PATH INTEGRAL APPROACH}
\end{center}

Consider a system where a macroscopic degree of freedom $R$ 
couples to an intrinsic degree of freedom $\xi$. 
We assume the following Hamiltonian for this system; 
\begin{equation}
H(R,\xi)=-{{\hbar^2}\over{2\mu}}{{\partial^2}\over{\partial R^2}} 
+U(R)+H_0(\xi)+V(R,\xi), 
\end{equation}
where $\mu$ is the mass for the macroscopic motion, 
$U(R)$ the potential in the absence of the coupling,  
$H_0(\xi)$ the Hamiltonian for the internal motion, and $V(R,\xi)$ 
the coupling between them.
The barrier transmission probability from the initial position $R_i$ 
on the right side of the barrier to the final position $R_f$ on the 
left side is then given by \cite{BT85}, 
\begin{eqnarray}
P(E)&=&\lim_{{R_i\rightarrow \infty}\atop{R_f\rightarrow -\infty}}
\biggl({P_i P_f \over {\mu^2}}\biggr)\int^\infty_0dT~e^{(i/\hbar)ET}  
\int^\infty_0 d{\widetilde T}~e^{-(i/\hbar)E{\widetilde T}} \nonumber \\
&\times & \int {\cal D}\bigl[R(t)\bigr] \int {\cal D}\bigl[ 
{\widetilde R}({\tilde t}) \bigr] e^{(i/\hbar)\lbrack S_t (R,T)-S_t ( 
{\widetilde R},{\widetilde T}) \rbrack}\rho_M \bigl({\widetilde R}
({\tilde t}),{\widetilde T};R(t),T\bigr), 
\end{eqnarray}
where $E$ is the total energy of the system,
and $P_i$ and $P_f$ are the classical momenta at $R_i$ and $R_f$, 
respectively. 
$S_t(R,T)$ is the action for the macroscopic motion along a path 
$R(t)$, and is given by 
\begin{equation}
S_t(R,T)=\int^T_0 dt \left(\frac{1}{2}\mu\dot{R}(t)^2 -U(R(t))\right). 
\end{equation}
The effects of the internal degree of freedom are included in the  
two time influence functional $\rho_M$, which is defined as 
\begin{equation}
\rho_M \bigl({\widetilde R}({\tilde t}),{\widetilde T};R(t),T\bigr) 
=\sum_{n_f}<n_i\vert \hat{u}^{\dagger}({\widetilde R}({\tilde t}),
{\widetilde T})\vert n_f><n_f\vert \hat{u}(R(t),T)\vert n_i> . 
\end{equation}
Here, $n_i$ and $n_f$ are the initial and the final 
states of the internal motion at position $R_i$ and $R_f$, respectively. 
The time evolution operator $\hat{u}(R(t),t)$ of the intrinsic motion 
obeys 
\begin{equation}
i\hbar \frac{\partial}{\partial t}\hat{u}(R(t),t)
=[H_0(\xi)+V(R(t),\xi)]\hat{u}(R(t),t). 
\end{equation}
with the initial condition $\hat{u}(R,t=0)=1$. 

We introduce the eigen-channel (or the adiabatic) basis, as
\begin{equation}
(H_0(\xi)+V(R,\xi))\varphi_n(R,\xi) 
=\lambda_n(R)\varphi_n(R,\xi). 
\end{equation}
Expanding the intrinsic wave function at time $t$ in this basis 
\begin{equation}
\hat{u}(R,t)|n_i> = \sum_n a_n(t)\exp\left[-\frac{i}{\hbar}
\int^t_0dt'\lambda_n\left(R(t')\right)\right]|\varphi_n(R(t))>,
\end{equation}
Eq. (2) for the barrier penetrability reads 
\begin{eqnarray}
P(E)&=&\lim_{{R_i\rightarrow \infty}\atop{R_f\rightarrow -\infty}}
\biggl({P_i P_f \over {\mu^2}}\biggr)\int^\infty_0dT~e^{(i/\hbar)ET}  
\int^\infty_0 d{\widetilde T}~e^{-(i/\hbar)E{\widetilde T}} \nonumber \\
&\times & \int {\cal D}\bigl[R(t)\bigr] 
\int {\cal D}\bigl[ {\widetilde R}({\tilde t}) \bigr] 
\sum_{n,m}a_n(T)a_m(\widetilde{T})^*
<\varphi_m({\widetilde R}({\tilde T}))|\varphi_n(R(T))> \nonumber \\
&\times & e^{\frac{i}{\hbar}\int^T_0dt\left(\frac{1}{2}\mu\dot{R}^2
-U(R)-\lambda_n(R)\right)}
e^{-\frac{i}{\hbar}\int^{\widetilde {T}}_0d{\tilde t}
\left(\frac{1}{2}\mu
\dot{{\widetilde {R}}}^2
-U(\widetilde{R})-\lambda_m(\widetilde{R})\right)}.
\end{eqnarray}
We use the semi-classical approximation, and for energies well below the 
barrier, where the single paht dominates, evaluate 
the path integral along the classical path 
\begin{equation}
R(t)=\widetilde{R}(\tilde{t})=R_{cl}(t),~~~~T=\widetilde{T}^*=T_{cl}. 
\end{equation}
In this case, the orthogonalitiy of the adiabatic basis leads to 
\begin{equation}
P(E)=\sum_n v_n(E)P_0(E; U(R)+\lambda_n(R)),
\end{equation}
where the weight factors are given by 
\begin{equation}
v_n(E)=|a_n(T_{cl})|^2. 
\end{equation}
The weight factors depend implicitly on the energy $E$ through 
the time evolution of the intrinsic system along the classical 
path.

\newpage

\begin{center}
{\bf Figure Captions}
\end{center}

\noindent
{\bf Fig. 1:} The eigen-potentials (the upper panel) and the associated 
weight factors (the lower panel) as a function of the relative 
distance $R$. The solid and the dashed lines in the lower panel are 
the weight factors for the lower and the higher eigen-potentials, 
respectively. 
The dotted line in the upper panel represents the bare potential barrier. 

\noindent
{\bf Fig. 2:} The penetrability (the first panel) and its first derivative 
(the second panel) for the two level problem as a function of the incident 
energy. The third panel shows the optimum weight factors obtained 
according to eq. (5). The solid and the dashed lines correspond to the 
weight factors for the lower and the higher potentials, respectively. 

\noindent
{\bf Fig. 3:} Same as fig. 2, but for different values of the 
excitation energy of the intrinsic motion which are denoted in the inset. 
The third panel shows only the weight factors for the lower potential. 

\noindent
{\bf Fig. 4:} 
Effects of the coupling to a high excitation energy state. 
The dotted line is the result without channel coupling, while the solid
line is obtained by taking into account the coupling to an excited
state whose excitation energy is 8 MeV. 

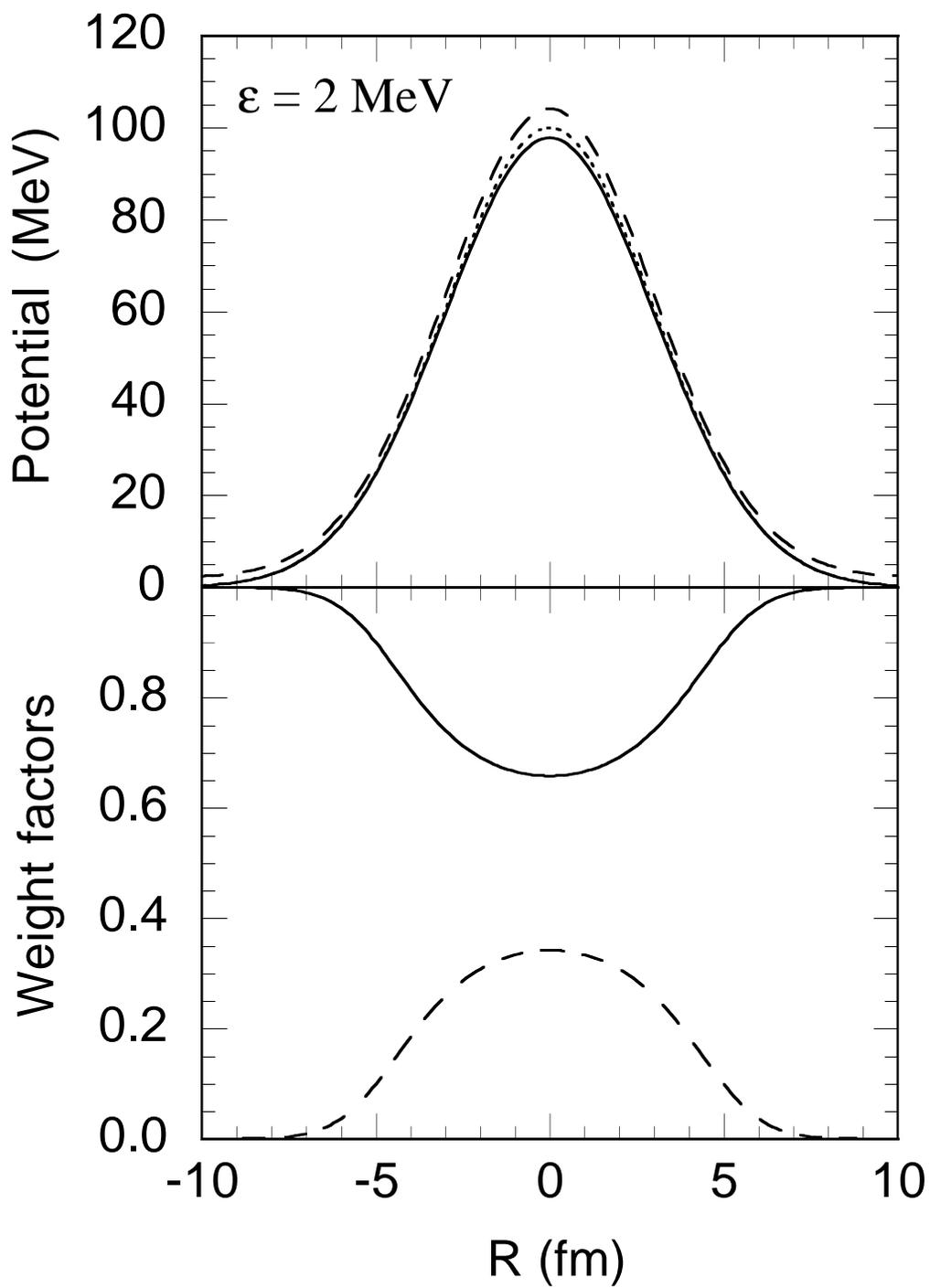
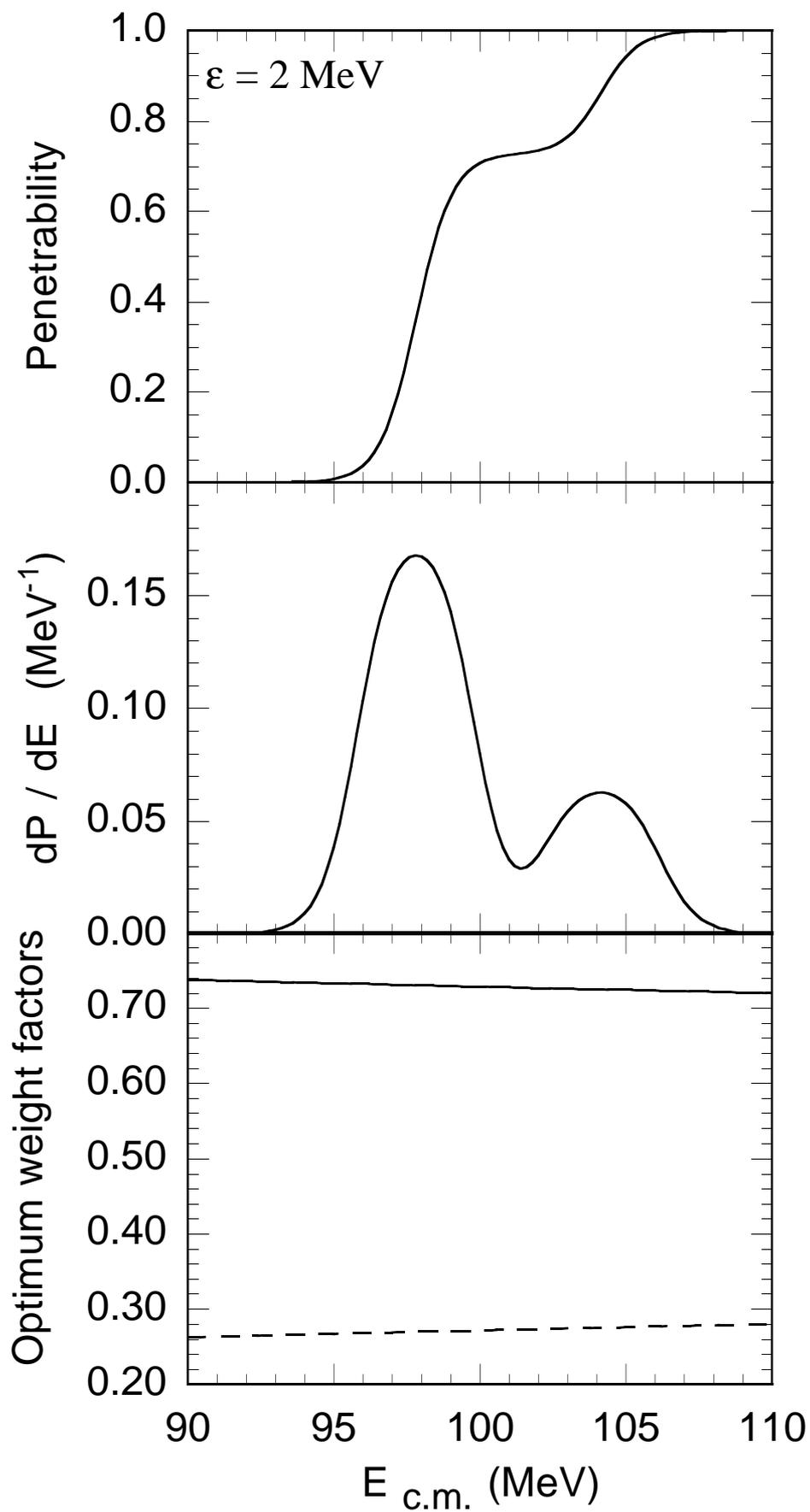
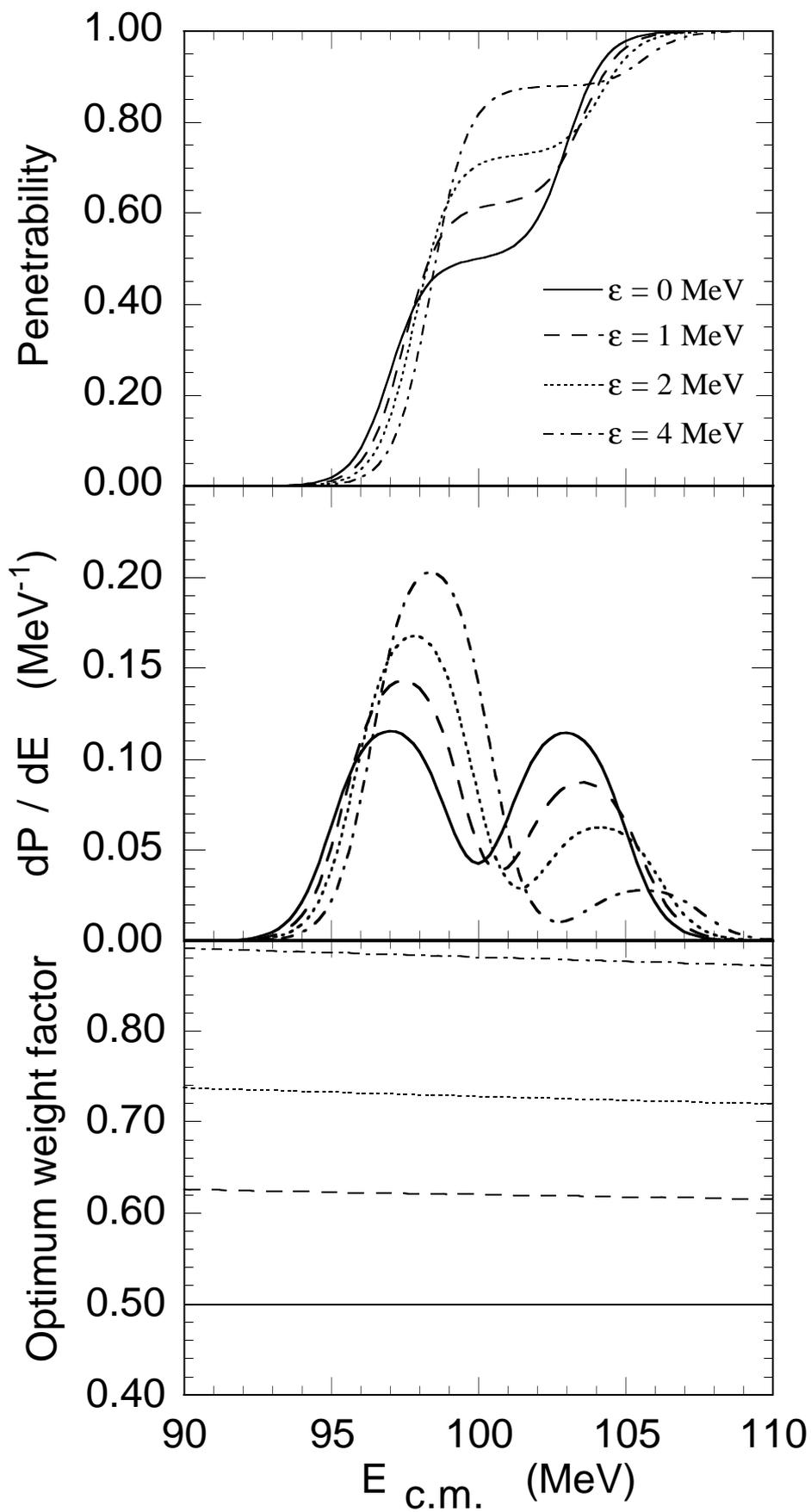
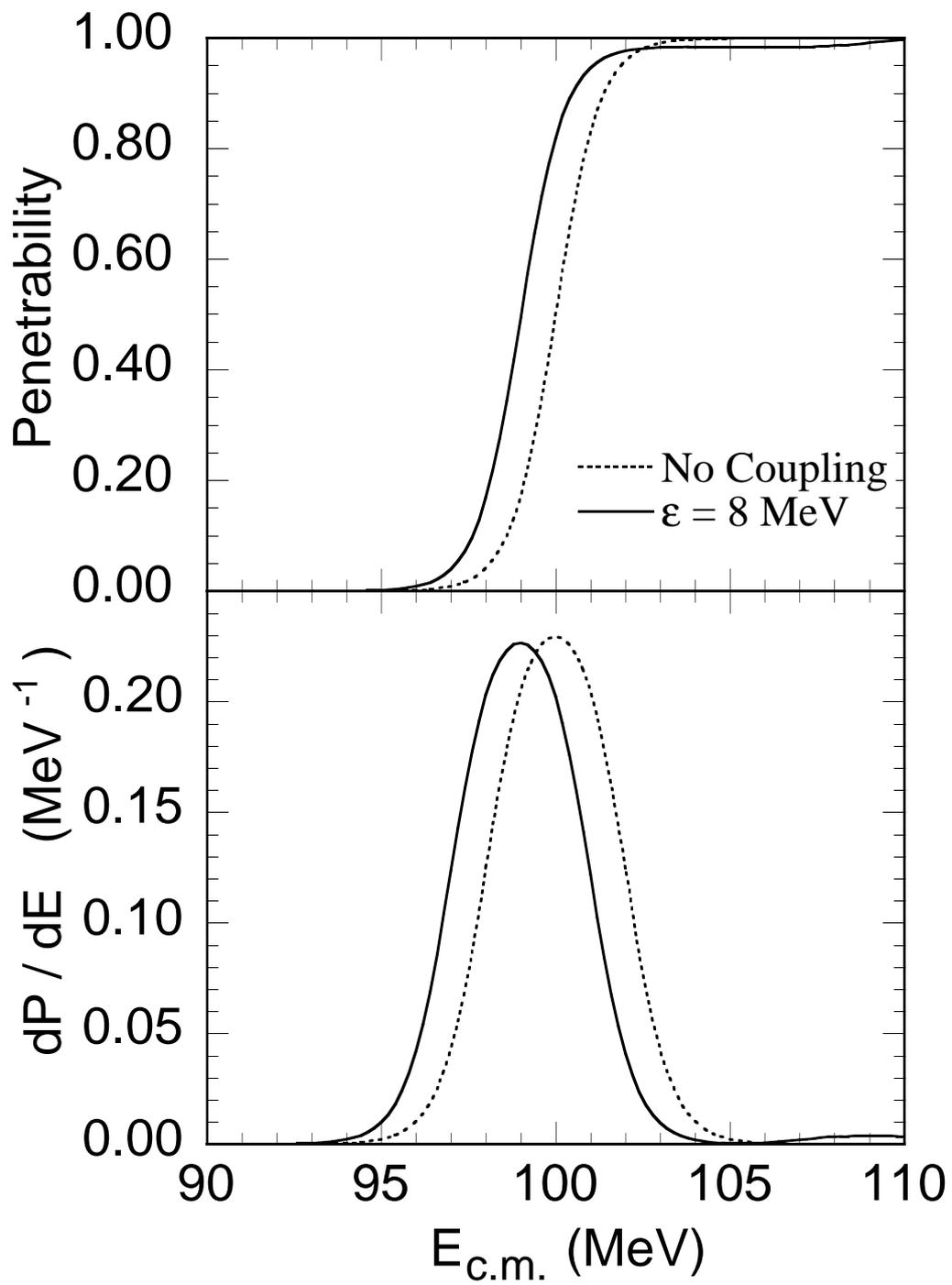
\end{document}